\documentclass[dvips]{PoS}
\usepackage{amsmath,amssymb}
\usepackage{graphicx}
\usepackage{subfigure}
\usepackage{units}

\title{Lattice investigation of the tetraquark candidates $a_0(980)$ and $\kappa$}

\ShortTitle{Lattice investigation of the tetraquark candidates $a_0(980)$ and $\kappa$}

\author{\speaker{Jan Oliver Daldrop}, Carsten Urbach\\
  Helmholtz-Institut f{\"u}r Strahlen und Kernphysik and
  Bethe Center for Theoretical Physics,\\
  Universit{\"a}t Bonn, Nussallee 14-16, 53115 Bonn, Germany\\
  E-mail: \email{daldrop,urbach@hiskp.uni-bonn.de}
}

\author{Constantia Alexandrou, Mario Gravina\\
Department of Physics, University of Cyprus, P.O.\ Box 20537, 1678 Nicosia, Cyprus\\
Computation-based Science and Technology Research Center, Cyprus Institute, \\ 20 Kavafi Str., Nicosia 2121, Cyprus
}

\author{Mattia Dalla Brida\\
School of Mathematics, Trinity College Dublin, Dublin 2, Ireland}

\author{Luigi Scorzato\\
ECT$^\star$, Strada delle Tabarelle, 286, I-38123, Trento, Italy
}

\author{Marc Wagner\\
Goethe-Universit\"at Frankfurt am Main, Institut f\"ur Theoretische Physik, \\ Max-von-Laue-Stra{\ss}e 1, D-60438 Frankfurt am Main, Germany

}

\author{for the European Twisted Mass Collaboration}

\abstract{
It is a long discussed issue whether light scalar mesons have sizeable
four-quark components. We present an exploratory study of this
question using $N_f=2+1+1$ twisted mass lattice QCD. A mixed action
approach ignoring disconnected contributions is used to calculate
correlator matrices consisting of mesonic molecule, diquark-antidiquark and two-meson interpolating operators with quantum
numbers of the scalar mesons $a_0(980)$ $(1(0^{++}))$  and $\kappa$
$(1/2(0^+))$. The correlation matrices are analyzed by solving the
generalized eigenvalue problem. The theoretically expected free
two-particle scattering states are identified, while no additional low
lying states are observed. We do not observe indications for bound
four-quark states in the channels investigated.
}

\FullConference{The 30th International Symposium on Lattice Field Theory\\
                 June 24 -- 29,  2012\\
                 Cairns, Australia}

\begin{document}

\section{Introduction}

In this project, we perform first steps towards extracting the
properties of the $a_0(980)$ and $\kappa$ resonances from lattice
QCD. 

The mass and width of an infinite volume resonance can theoretically
be extracted from Euclidean lattice QCD by studying the corresponding
energy levels as a function of the lattice size $L$ with an extension
of L\"uscher's finite volume method
\cite{Luscher:1985dn,Luscher:1986pf,Luscher:1990ck,Luscher:1990ux,Luscher:1991cf}.

However, such an analysis is challenging because several energy levels
have to be extracted from the lattice simulation with high accuracy.
Therefore, we follow an exploratory approach: we perform a study
of correlator matrices of four-quark operators in order to investigate
whether bound four-quark states are observed in the $a_0(980)$ or
$\kappa$ channels. Recently, in such a set-up in $N_f=2$ QCD, hints at
bound tetraquark states in the $\sigma$ and $\kappa$ channels have
been identified~\cite{Prelovsek:2010kg}. 

We compute the correlator matrices in a mixed
action \cite{Baron:2010th,Baron:2010vp} analysis of four-quark states
on the lattice ignoring disconnected contributions. The goal is to
answer the question whether there is a 
bound tetraquark or molecule state in the $a_0(980)$ or
$\kappa$ channel. Identifying such a state in addition to the free
scattering states could be a hint on the nature of the corresponding
resonance. Additionally, the overlap to different interpolating operators
could be studied.

The lattice study is based on $N_f=2+1+1$ gauge configurations generated by the European Twisted Mass (ETM) Collaboration. 
Details on the generation of the ensembles can be found in Ref.~\cite{Baron:2010bv}.
The analysis is performed on four ensembles with a lattice spacing of $a\approx 0.086\,\unit{fm}$ and pion masses between $m_{\pi^+}\approx 280 \,\unit{MeV}$ and $m_{\pi^+}\approx 450\,\unit{MeV}$, see Table~\ref{tab:setup}.

For the computation of observables we use a twisted mass discretization for the valence $s$ quarks, which is different from the sea $s$ quarks to avoid the problem of mixing between $s$ and $c$ quarks, for details cf.\ Refs.\ \cite{Baron:2010th,Baron:2010vp}.

\section{Interpolating operators and extraction of energy levels}

To perform a four-quark analysis of $a_0(980)$, we compute
correlators $C_{ij}(t) = < O_i(t) O_j^\dagger(0) >$ with interpolating
operators

\begin{eqnarray}
 O_i&=&\sum_{\boldsymbol{x}}\sum_{\mu=1,2,3}\left( \bar{d}(\boldsymbol{x}) \Gamma_i^\mu s(\boldsymbol{x})\right)\left(\bar{s}(\boldsymbol{x})\Gamma_i^\mu u(\boldsymbol{x})\right) \,\,\,\text{for}\,\,\,i=1,2,3, \label{eqn:o123}\\
 O_i&=&\sum_{\boldsymbol{x}}\left[\bar{d}(\boldsymbol{x}){\Gamma_i}\bar{s}^T(\boldsymbol{x})\right]_a\,\left[s^T(\boldsymbol{x})\Gamma_i u(\boldsymbol{x})\right]_a  \,\,\,\text{for}\,\,\,i=4,5,\label{eqn:o45} \\
 O_6&=&\sum_{\boldsymbol{x}}\left(\bar{s}(\boldsymbol{x})\gamma_5 s(\boldsymbol{x})\right) \left(\bar{d}(\boldsymbol{x}) \gamma_5 u(\boldsymbol{x})\right),
\label{eqn:o6}
\end{eqnarray}
with the $\gamma$-matrices
$
 \Gamma_1=\gamma_5,\,\,\,\Gamma_2^\mu=\gamma^\mu,\,\,\, \Gamma_3^\mu=\gamma^\mu\gamma_5,\,\,\,\Gamma_4=C\gamma_5,\,\,\,\Gamma_5=C,\,\,\,C=\gamma_0\gamma_2.
$
Due to the omission of disconnected diagrams the artificial pseudoscalar $\bar{s}\gamma_5 s$ state, which we call $\eta_s$ in the following, will be relevant for the calculation. To ensure the correct identification of the $\eta_s\pi$ state, we explicitly included the operator $O_6$. \\ 
For further clarification of the nature of the examined states, the two-meson operators
\begin{eqnarray}
 O_7&=&\sum_{\boldsymbol{x}}\left(\bar{d}(\boldsymbol{x})\gamma_5 s(\boldsymbol{x})\right)\sum_{\boldsymbol{y}} \left(\bar{s}(\boldsymbol{y}) \gamma_5 u(\boldsymbol{y})\right), \label{eqn:o7} \\
 O_8&=&\sum_{\boldsymbol{x}}\left(\bar{s}(\boldsymbol{x})\gamma_5 s(\boldsymbol{x})\right)\sum_{\boldsymbol{y}} \left(\bar{d}(\boldsymbol{y}) \gamma_5 u(\boldsymbol{y})\right) \label{eqn:o8}
\end{eqnarray}
are studied as well (on a single ensemble).

In order to extract energy levels from the correlator matrices, we
solve the generalized eigenvalue problem (GEVP)
\cite{Luscher:1990ck,Michael:1982gb,Blossier:2009kd}.  
However, due to the
presence of pairs of pseudoscalar meson states $| M_1 >$
and $| M_2>$ coupling to our operators according to
$<M_1 | O_j | M_2> \neq 0$
problems arise: one of the two mesons can travel forward the other
backward in time, severely complicating the analysis of our correlators. To
avoid these problems, we restrict the analysis to $t
\lesssim T/4$ and $T-t \lesssim T/4$, because the single-meson contributions are not relevant for sufficiently small
$t$ (or $T-t$). For a more detailed
discussion we refer to an upcoming publication and to the Refs.\ \cite{Prelovsek:2010kg, Detmold:2008yn, Prelovsek:2008rf}. Excited state contributions
are taken into account by performing two-mass fits to the eigenvalues
of interest.

\begin{table}[t!]
  \centering
  \begin{tabular*}{.8\textwidth}{@{\extracolsep{\fill}}lccccccc }
    \hline\hline
    ensemble & $\beta$ & $a\mu_\ell$ & $a\mu_\sigma$ & $a\mu_\delta$ & 
    $L/a$ & $N_\text{conf}$ & smearing \\
    \hline\hline
    A30.32  & $1.90$ & $0.0030$ & $0.150$ & $0.190$ & $32$ & 672 & APE \\
    A40.20  & $1.90$ & $0.0040$ & $0.150$ & $0.190$ & $20$ & 500 & none \\
    A40.24  & $1.90$ & $0.0040$ & $0.150$ & $0.190$ & $24$ & 1259 & APE\\
    A80.24  & $1.90$ & $0.0080$ & $0.150$ & $0.190$ & $24$ & 1225 & APE\\
    \hline\hline
  \end{tabular*}
  \caption{The input parameters of the ensembles used in this project, the number of configurations and the smearing type employed.}
  \label{tab:setup}
\end{table}

\section{Results}

\subsection{$a_0(980)$: four-quark and two-particle operators, a single ensemble}

We start by discussing $a_0(980)$ ($I(J^{PC}) = 1(0^{++})$) results
obtained using ensemble A40.20 (cf.\ Table~\ref{tab:setup}). This
ensemble with rather small spatial extent ($L \approx 1.72 \,
\textrm{fm}$) is particularly suited to distinguish two-particle
states with relative momentum from states with two particles at rest
and from possibly existing $a_0(980)$ four-quark states (the former
have a rather large energy because one quantum of momentum is
$p_\textrm{min} = 2 \pi / L \approx 720 \, \textrm{MeV}$). 

\begin{figure}[htb]
\begin{center}
\begin{tabular}{cc}
\text{(a)} & \text{(b)} \\
\includegraphics[width=0.45\textwidth]{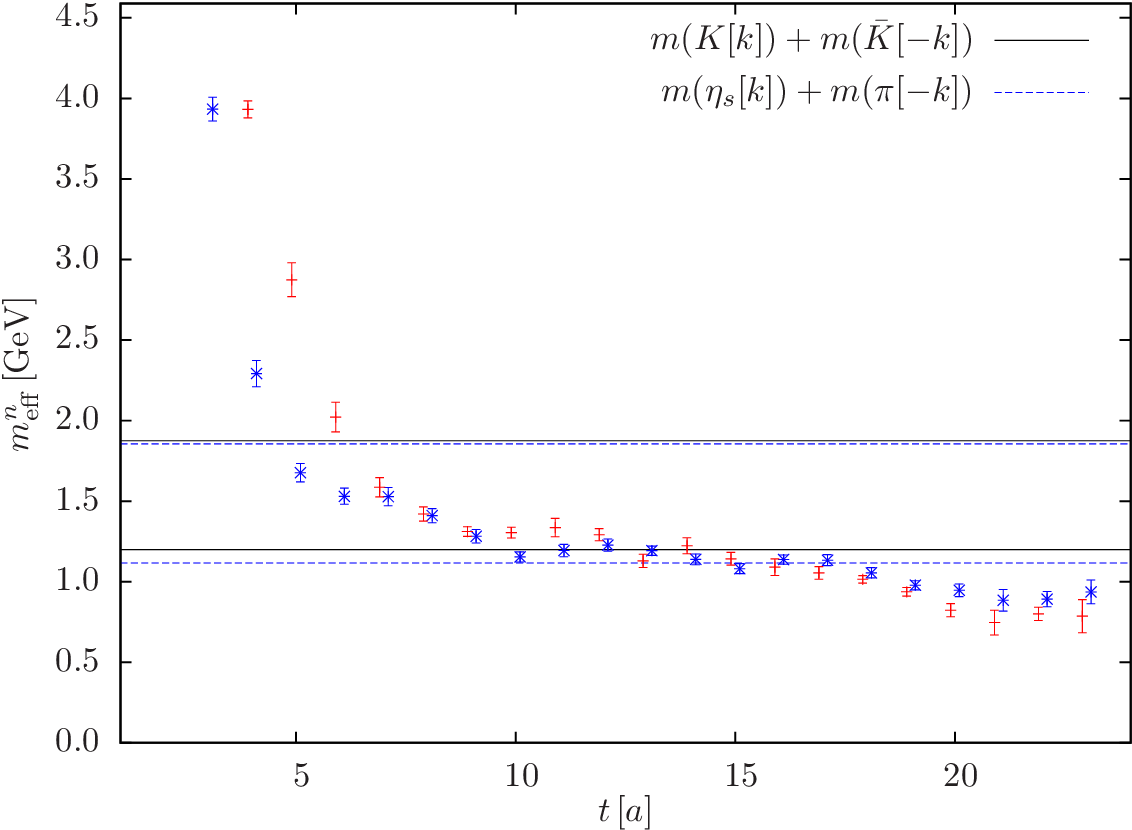} &
\includegraphics[width=0.45\textwidth]{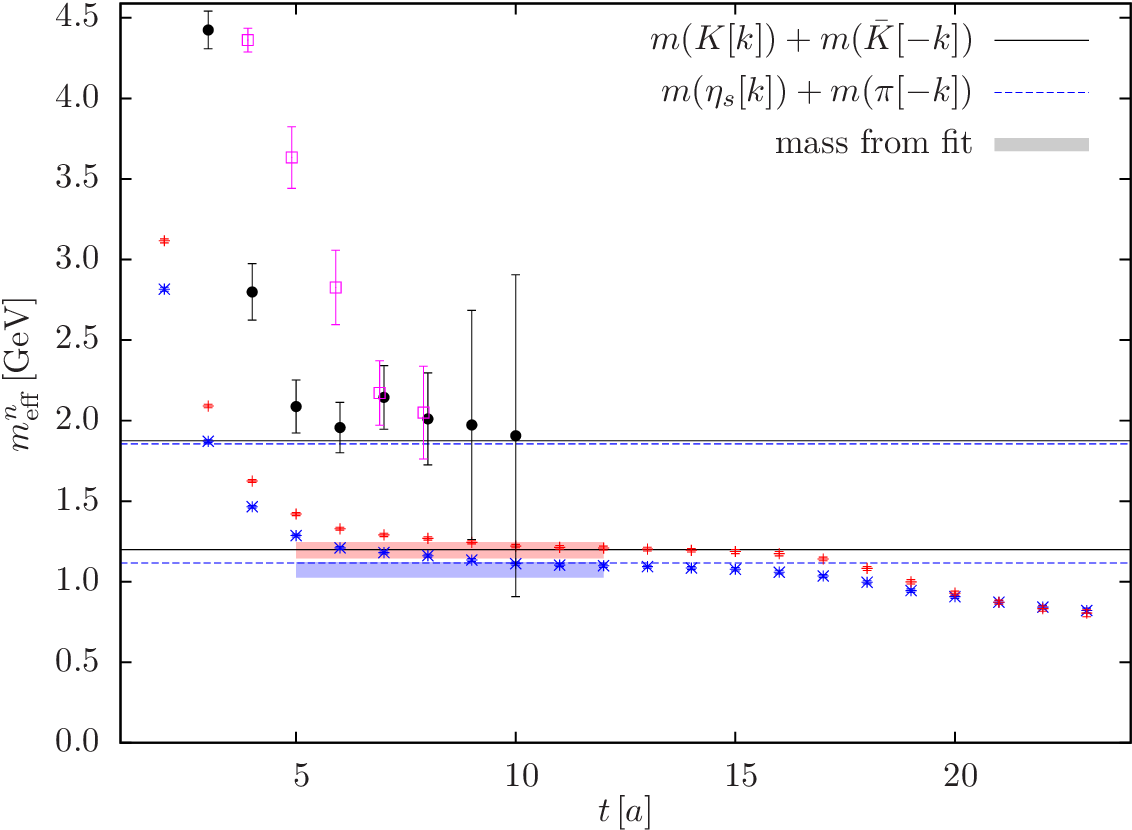} \\

 & \\
\text{(c)} & \text{(d)} \\
\includegraphics[width=0.45\textwidth]{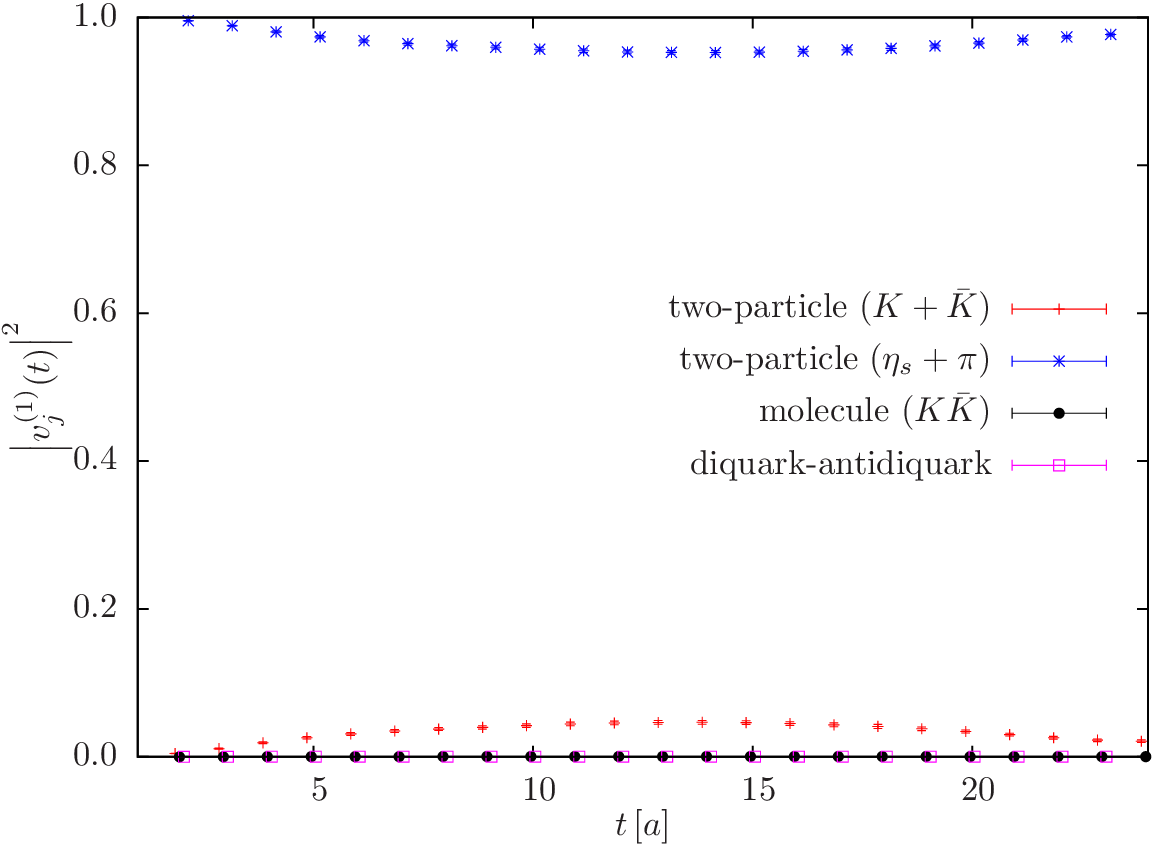}  &
\includegraphics[width=0.45\textwidth]{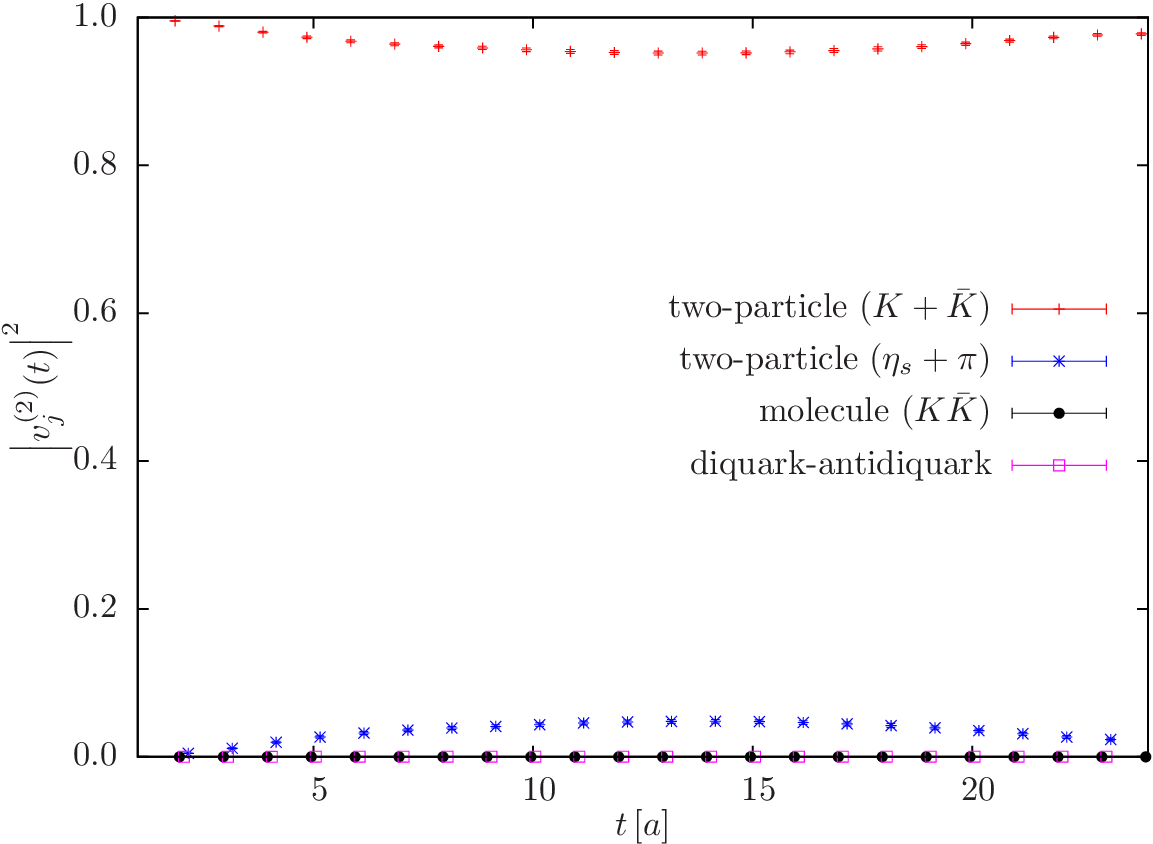}
\end{tabular}
\caption{\label{FIG001}$a_0(980)$ sector, A40.20 ensemble.
\textbf{(a)}~Effective masses as functions of the temporal separation, $2 \times 2$ correlation matrix (local operators: $K \bar{K}$ molecule, diquark-antidiquark). Horizontal lines indicate the expected two-particle $K + \bar{K}$ and $\eta_s + \pi$ energy levels.
\textbf{(b)}~$4 \times 4$ correlation matrix (local operators: $K \bar{K}$ molecule, diquark-antidiquark, two-particle $K + \bar{K}$, two-particle $\eta + \pi$).
\textbf{(c)}, \textbf{(d)}~Squared eigenvector components of the two low-lying states from \textbf{(b)} as functions of the temporal separation.}
\end{center}
\end{figure}

Figure~\ref{FIG001}a shows effective mass plots from a $2 \times 2$
correlation matrix with a $K \bar{K}$ molecule operator $O_1$ (see equation
(\ref{eqn:o123})) and a diquark-antidiquark operator $O_4$
(\ref{eqn:o45}). The corresponding two plateaus are around $1000 \,
\textrm{MeV}$ and, therefore, consistent both with a possibly existing
$a_0(980)$ four-quark state and with two-particle $K + \bar{K}$ and
$\eta_s + \pi$ states, where both particles are at rest. 

Increasing this correlation matrix to $4 \times 4$ by adding two-particle $K + \bar{K}$ and $\eta_s + \pi$ operators (equations\ (\ref{eqn:o7}) and (\ref{eqn:o8})) yields the effective mass plot shown in Figure~\ref{FIG001}b. The same two low-lying states are resolved, however, with significantly
better quality. Two additional states are observed, whose plateaus are around $1500 \, \textrm{MeV} - 2000 \, \textrm{MeV}$. From this $4 \times 4$ analysis we conclude the following:

1.\ We do not observe a third low-lying state around $1000 \, \textrm{MeV}$, even though we provide operators, which are of four-quark type as well as of two-particle type. This suggests that the two low-lying states are the expected two-particle $K + \bar{K}$ and $\eta_s + \pi$ states, while no additional stable $a_0(980)$ four-quark state does exist in the A40.20 ensemble.

2.\ The effective masses of the two low-lying states are of much better quality in Figure~\ref{FIG001}b than in Figure~\ref{FIG001}a. We attribute this to the two-particle $K + \bar{K}$ and $\eta_s + \pi$ operators, which presumably create larger overlap to those states than the four-quark operators. This in turn confirms the interpretation of the two low-lying states as two-particle states.

3.\ To investigate the overlap in a more quantitative way, we show the squared eigenvector components of the two low-lying states in Figure~\ref{FIG001}c and Figure~\ref{FIG001}d (cf.\ Ref.\ \cite{Baron:2010vp} for a more detailed discussion of such eigenvector components). Clearly, the lowest state is of $\eta_s + \pi$ type, whereas the second lowest state is of $K + \bar{K}$ type. On the other hand, the two four-quark operators are essentially irrelevant for resolving those states. These eigenvector plots give additional strong support of the above interpretation of the two low lying states as two-particle states.

4.\ The estimated energy levels of two-particle excitations with one relative quantum of momentum are consistent with the effective mass plateaus of the second and third excitation in Figure~\ref{FIG001}b.

$ $ \\

Figure~\ref{FIG001}a and Figure~\ref{FIG001}b also demonstrate that two-particle states can be resolved by four-quark operators, i.e.\ two-particle operators are not necessarily needed, to extract the full spectrum. Since we are mainly interested in possibly existing states with a strong four-quark component, we restrict the correlation matrices computed for other ensembles to four-quark operators.

\subsection{$a_0(980)$: four-quark operators, many ensembles}

We have analyzed the three additional ensembles listed in Table~\ref{tab:setup} with respect to $a_0(980)$ in a similar way as explained in the previous subsection.
The main difference is that this time we exclusively use four-quark operators, but no two-particle operators. To be able to resolve more than two low-lying states, we use the operators $O_1$ to $O_6$.

\begin{figure}[htb]
\centering
\begin{tabular}{cc}
 $a_0(980)$ {sector} & $\kappa$ {sector}\\
\includegraphics[width=0.45\textwidth]{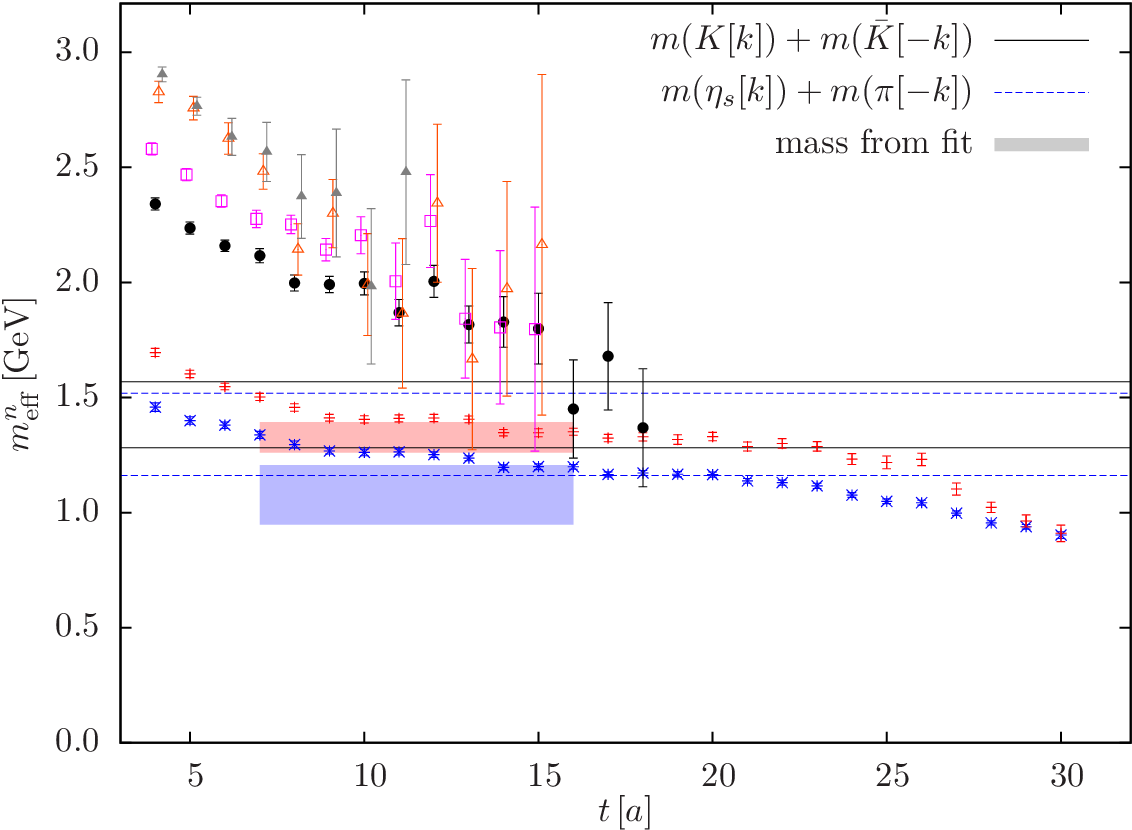} &
\includegraphics[width=0.45\textwidth]{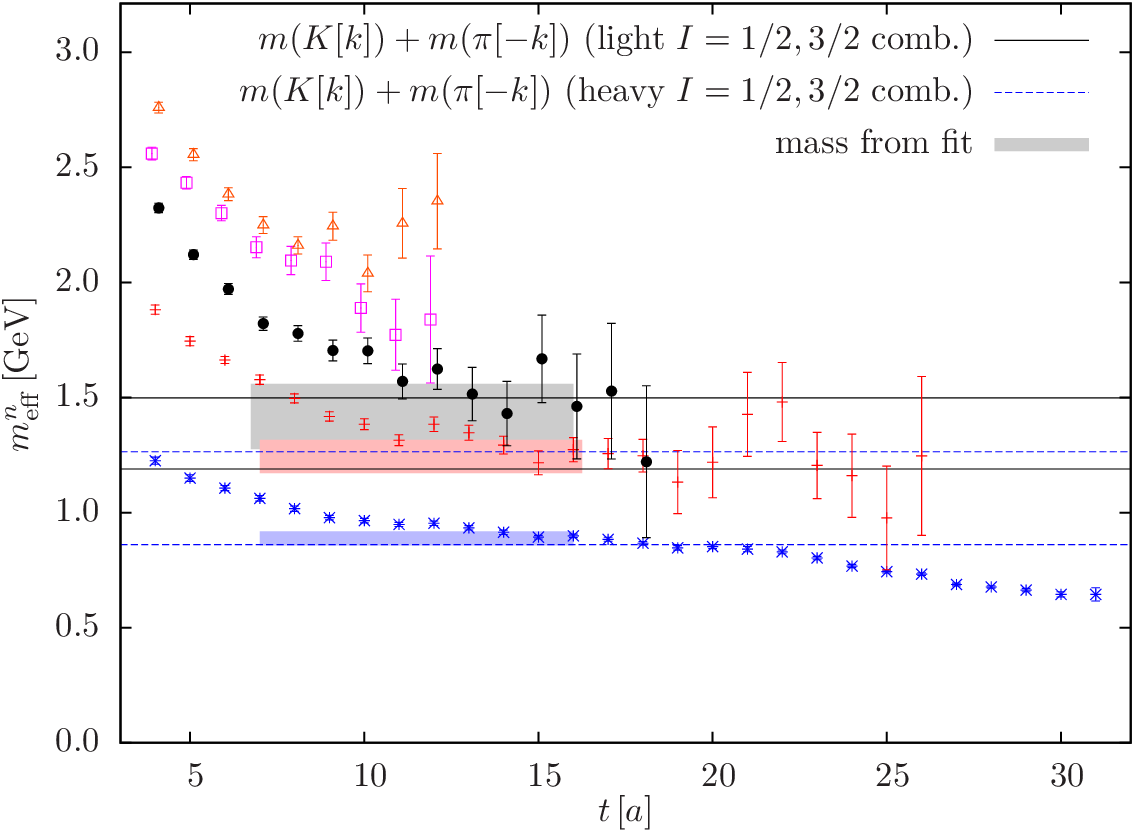} \\

\end{tabular}

\caption{\label{FIG002} Effective masses as functions of the temporal separation for the $a_0(980)$ and $\kappa$ sector for the A30.32 ensemble. Horizontal lines indicate the expected two-particle $K + \bar{K}$ and $\eta_s + \pi$ energy levels. Additionally, the masses extracted from fits are included.}

\end{figure}

An effective mass plot for the A30.32 ensemble (cf.\
Table~\ref{tab:setup}) is
shown in Figure~\ref{FIG002} together with the expected energy levels of the relevant two-particle
states and the masses extracted by fits. The effective mass plots for the other ensembles will be shown in an upcoming publication. 

On a qualitative level our findings agree for all ensembles, i.e.\ are
as reported in the previous subsection: there are always two
low-lying states, whose masses are consistent with the expected masses
of the two-particle $K + \bar{K}$ and $\eta_s + \pi$ states; higher
excitations (the third, forth, etc.\ extracted state) are in all cases
significantly heavier and consistent with two-particle excitations
with one relative quantum of momentum.

\subsection{\label{SEC001}$\kappa$: four-quark operators, many ensembles}

The analysis for the $\kappa$ sector ($I(J^P) = 1/2 (0^+)$) closely parallels the analysis of the $a_0(980)$ sector presented above.
We consider $5 \times 5$ correlation matrices containing a $K \pi$ molecule operator analogue to $O_1$ and further operators corresponding to $O_2$ to $O_5$ with the appropriate valence quark content, so that our operators are essentially identical to those considered in Ref.\ \cite{Prelovsek:2010kg}.

In twisted mass lattice QCD the isospin $I$ is not a quantum number. Therefore, it is not sufficient to only resolve $I = 1/2$ two-particle $K + \pi$ states. One has to take into account also mixing with $I = 3/2$ two-particle $K + \pi$ states, i.e.\ it is necessary to resolve these low-lying two-particle states at the same time (details will be discussed in an upcoming publication).

An effective mass plot for the A30.32 ensemble is shown in Figure~\ref{FIG002}
together with the expected energy levels of two-particle $K + \pi$
states and the masses extracted by fits. While effective mass plateaus
are consistent with these expected two-particle energy levels, there
is no indication of any additional low lying state, i.e.\ of a
possibly existing bound four-quark $\kappa$ state. While this is suggested by
experimental data, it contradicts the findings of the similar recent
lattice study of $\kappa$ \cite{Prelovsek:2010kg}.

\section{Summary and Outlook}

We computed the low-lying spectrum in the $a_0(980)$ and $\kappa$
sectors by employing trial states designed to have a substantial
overlap with both two-particle and possibly existing tetraquark
states. With our ensembles, we did not see additional states beside
those that can be identified with the expected two-particle
spectrum. The next states appear roughly consistent with excitations
of the first quantum of momentum 
($2\pi/L$) on top of those thresholds. This is somewhat difficult to
reconcile with the additional state found in Ref.\
\cite{Prelovsek:2010kg} in the $\kappa$ channel, despite the rather
similar lattice setups. 

We find that the low lying spectrum has essentially exclusively overlap to two-particle trial states.  This
suggests that the states that we see are, indeed, the expected two-particles states at the threshold and not
tightly bound states either of molecular type or diquark-antidiquark type.

These conclusions can be strengthened by studying more volumes, by introducing twisted boundary
conditions \cite{Bedaque:2004kc} and by studying further trial states of different type.  As for the latter, it will be crucial to
combine four quarks with traditional quark-antiquark operators
including disconnected diagrams. 
As for the volume dependence, we plan to use the finite volume formulae of L\"uscher \cite{Luscher:1985dn,
Luscher:1986pf, Luscher:1990ck, Luscher:1990ux, Luscher:1991cf} and their extensions to multiple channels developed
in Refs.\ \cite{Lage:2009zv,Bernard:2010fp,Oset:2011ce,Doring:2011vk}. At present, our limited number of volumes is
insufficient for such an analysis. Corresponding computations are in progress.

\subsection*{Acknowledgments}

It is a great pleasure to thank Akaki Rusetsky for many enlightening discussions. We also acknowledge helpful discussions with Vladimir Galkin and Vincent Mathieu. We thank Konstantin Ottnad for providing analysis code. Furthermore, we like to thank the J\"ulich Supercomputing Centre (JSC) and AuroraScience for support and computation time, and we are grateful for the support by the ETM Collaboration and by the DFG via
the Sino-German CRC 110. M.G. was supported by the Marie-Curie European training network ITN STRONGnet grant PITN-GA-2009-238353. M.D.B. warmly thanks the support and hospitality of the AuroraScience project, ECT*, and the University of Cyprus, where part of this work was carried out.
M.W.\ acknowledges support by the Emmy Noether Programme of the DFG
(German Research Foundation), grant WA \mbox{3000/1-1}. This work was supported
in part by the Helmholtz International Center for FAIR within the
framework of the LOEWE program launched by the State of Hesse.

\end{document}